\documentclass[preprint,nofootinbib,showpacs,12pt,floatfix]{revtex4-1}

\usepackage{amssymb}
\usepackage{graphicx}
\usepackage{dcolumn}
\usepackage{bm}
\usepackage{color}
\usepackage{latexsym}
\usepackage{lineno}
\usepackage{comment}
\usepackage{color}
\usepackage{marginnote}

\begin{document}
\newcommand{\leftg}{\langle \phi_0 |}
\newcommand{\rightg}{| \phi_0 \rangle}
\newcommand{\chiral}{\langle \bar{q} q \rangle} 
\newcommand{\vs}{\vspace{-0.25cm}}
\newcommand{\gev}{\,\mathrm{GeV}}
\newcommand{\mev}{\,\mathrm{MeV}}
\newcommand{\fmd}{\,\mathrm{fm}^{-3}}
\newcommand{\ord}[1]{\mathcal{O}(k_f^{#1})}
\newcommand{\dif}{\mathrm{d}}

\title{Nuclear pairing from microscopic forces:
singlet channels and higher-partial waves
}
\author{Stefano Maurizio}
\affiliation{Department of Physics, University of Bologna, I-40126 Bologna, Italy}
\author{Jeremy W. Holt}
\affiliation{Department of Physics, University of Washington, Seattle, WA 98195-1560}
\author{Paolo Finelli}
\affiliation{Department of Physics, University of Bologna, I-40126 Bologna, Italy\\
INFN, Bologna section, I-40127 Bologna, Italy}

%
%

\begin{abstract}
{\bf Background:} An accurate description of nuclear pairing gaps is extremely important 
for understanding static and dynamic properties of the inner crusts of neutron stars 
and 
to explain their cooling process.

{\bf Purpose:} We plan to study the behavior of the pairing gaps $\Delta_F$ as a function of 
the Fermi momentum $k_F$ for neutron and nuclear matter in all relevant angular momentum 
channels where superfluidity is believed to naturally emerge. The calculations
will employ realistic chiral nucleon-nucleon potentials 
with the inclusion of three-body forces and self-energy effects. 

{\bf Methods:}
The superfluid states of neutron and nuclear matter are studied by solving the BCS gap 
equation 
for chiral nuclear potentials using the method suggested by Khodel {\it et al.}, 
where the original gap equation 
is replaced by a coupled set of equations for the dimensionless gap function $\chi(p)$ defined 
by $\Delta(p) =  \Delta_F \chi(p)$ and a non-linear
algebraic equation for the gap magnitude  $\Delta_F = \Delta(p_F)$ at the Fermi surface. This
method is numerically stable even for small pairing gaps, such as that encountered in the 
coupled $^3PF_2$ partial wave.

{\bf Results:} We have successfully applied Khodel's method to singlet ($S$) and coupled channel 
($SD$ and $PF$) cases in neutron and nuclear matter. Our calculations agree with 
other {\it ab-initio} approaches, where available, and provide crucial inputs for future 
applications 
in superfluid systems.

\end{abstract}
\pacs{21.30.-x; 21.65.-f; 26.60.-c}
\maketitle

%
%

\section{Introduction}

Superfluidity in neutron matter is connected to different aspects of 
neutron star physics. At the surface of the star \cite{review_superfluidity,Chamel_Haensel_Rev}, where a neutron 
gas moves in a lattice structure of neutron-rich nuclei and a sea of relativistic electrons, 
a $^{1}S_0$ neutron pairing gap naturally emerges, while at larger densities a (possibly 
anisotropic) $^3PF_2$ gap plays a more important role (in particular for neutron star cooling 
\cite{PhysRevLett.106.081101, Cooling_Rev}). At the same time, the nuclear matter case
could be interesting for finite nuclear systems where neutron-proton pairing is relevant 
\cite{Jensen}, even if the appearance of pairing in ordinary uniform matter is probably 
questionable because of known instabilities \cite{RavenhallLattimer} which could hide 
superfluidity in a broad range of densities.

The goal of this article is to solve the BCS equations starting from modern
nucleon-nucleon (NN) forces based on chiral effective field theory \cite{Weinberg,Machleidt:2011zz,
Epelbaum:2008ga}. In this approach one identifies the appropriate low-energy degrees 
of freedom and derives the most general Lagrangian compatible with the symmetries and 
symmetry-breaking pattern of the underlying fundamental theory (i.e., QCD). 
The first steps towards a realistic NN potential from first principles started almost twenty years 
ago within the framework of Chiral Perturbation Theory (ChPT) \cite{Ordonez92,BKM}. In ChPT 
the nuclear potential emerges naturally as a hierarchy of terms controlled by a power 
expansion in $Q/\Lambda_\chi$, where $Q$ is a soft scale (pion mass, nucleon 
momentum) and $\Lambda_\chi$ is a hard scale (the nucleon mass $M_N$ or the chiral 
symmetry breaking scale $4\pi f_\pi$). Two-nucleon forces appear at leading order 
$(Q/\Lambda_\chi)^0$, while three-nucleon forces appear first at order $(Q/\Lambda_\chi)^3$, 
or next-to-next-to-leading order (N2LO). 

We employ primarily the high-precision NN potential developed in Ref.\ \cite{Machleidt:2011zz}
at next-to-next-to-next-to-leading order (N3LO) in the chiral expansion, but to asses theoretical
uncertainties associated with the choice of cutoff scale and regulating functions, we employ 
in addition the chiral nuclear potentials developed in Ref.\ \cite{Epelbaum:2008ga} in selected 
cases. To implement the leading three-nucleon force, we include a two-body density-dependent 
potential \cite{Holt:2009, Holt:2010} (see also Refs.\ \cite{Hebeler:2009iv,Lesinski:2011rn,Gandolfi,
Lovato:2010ef} for other approaches and relevant details). To improve convergence in 
many-body perturbation theory, it is desirable to employ nuclear interactions with a cutoff scale below
$\Lambda \sim 500$\,MeV. One approach is to employ renormalization group (RG) methods 
that decouple the low- and high-momentum components of the potential. Two 
different methods for evolving nuclear potentials to block- and band-diagonal form in a
momentum-space representation, $V_{{\rm lowk}}$ and $V_{{\rm srg}}$ respectively, have been
developed (see Refs.\ \cite{bogner03,rengroup,srgweb} 
for detailed reviews) and used in the present study. 
An alternative approach would be to construct from the beginning chiral 
nuclear potentials at lower cutoff scales \cite{Coraggio07,Coraggio13,Coraggio14}.

The paper is organized as follows. Section \ref{secII} introduces the BCS theory that is the 
standard framework for a microscopic description of nucleonic pairing. In 
particular, the numerical implementation first introduced by Khodel {\it et al.} \cite{Khodel S} will be reviewed.
Sections \ref{1s0_sec} and \ref{hpw_sec} describe, respectively, our predictions for pairing 
gaps in the singlet and in the coupled channel cases. The role of the two-body NN interaction 
will be discussed along with the influence of three-body forces and self-energy effects. 
Finally, Section \ref{secIV} presents our conclusions. 

%
%

\section{The BCS equation}
\label{secII}

\begin{figure}
\begin{center}
\includegraphics[width = 12.5 cm]{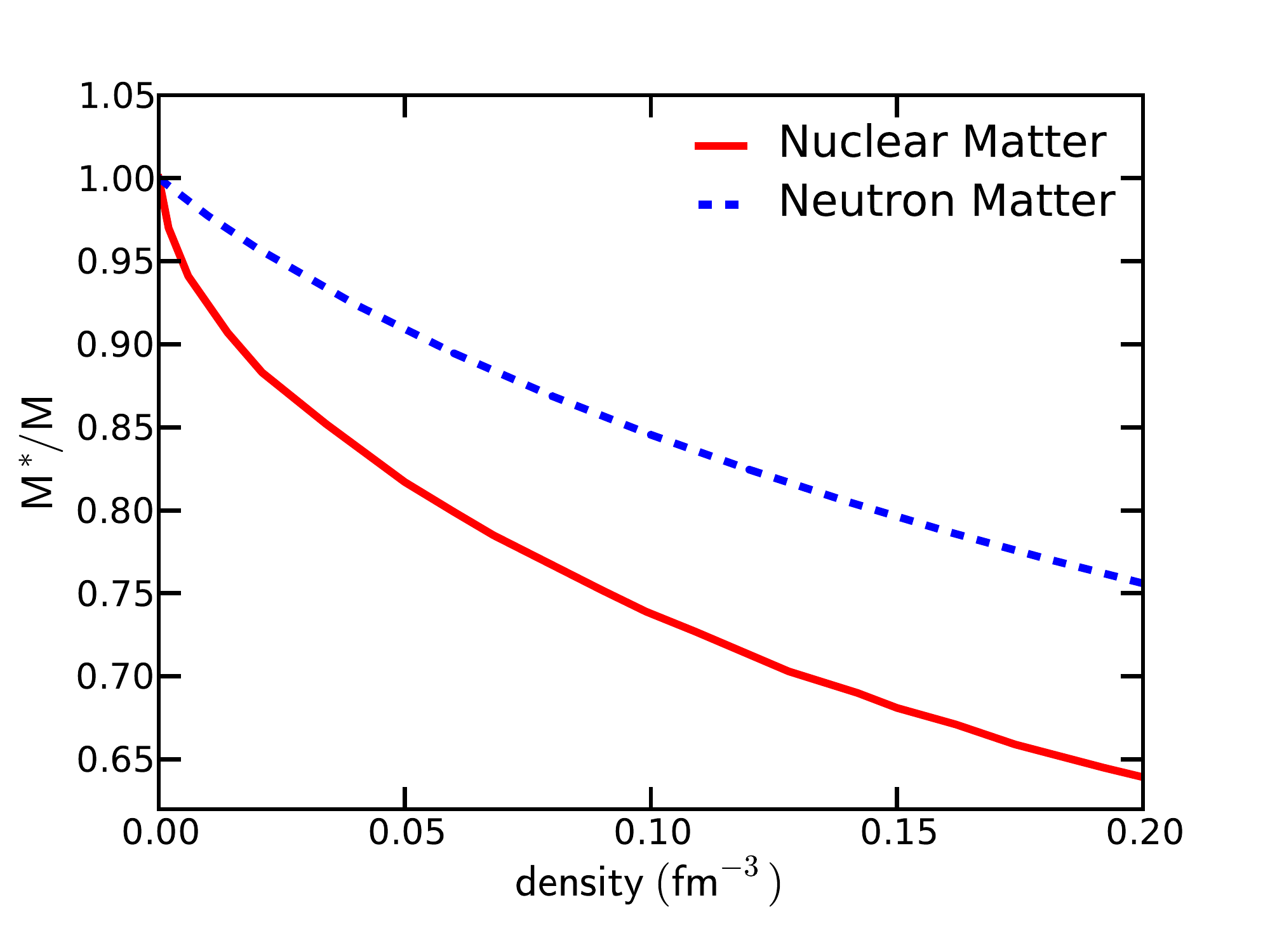}
\caption{(color online) The effective mass in the case of nuclear (red line) and neutron (blue dashed line) matter 
as a function, respectively, of the total nucleon density $\rho$ or the neutron density $\rho_n$ 
(see Ref.\ \cite{meff1}) .
}\label{fig:meff}
\end{center}
\end{figure}

In this section we explain the method employed to solve the BCS equations \cite{bcs} by 
partial-wave decomposition \cite{TaTa, Khodel S, Khodel, Jensen}. For simplicity
we largely neglect spin and isospin degrees of freedom in the derivation.
The BCS equation reads in terms of the NN potential $ V(\textbf{k}, \textbf{k}')  =
\langle{ \textbf{k} \left| V\right|  \textbf{k}'  }\rangle$ as follows
\begin{equation}
\Delta \left(\textbf{k}\right) = -\sum_{\textbf{k}'} \langle{ \textbf{k} \left| V\right|  \textbf{k}'  }
\rangle \frac{ \Delta \left(\textbf{k}'\right)}{ 2 E\left( \textbf{k}' \right)} \; ,
\label{gapeq}
\end{equation}
with $ E(\textbf{k})^2=\xi(\textbf{k})^2+\left|\Delta(\textbf{k})\right|^2$ and where 
$\xi(\textbf{k})=\varepsilon(\textbf{k})-\mu$, $\varepsilon(\textbf{k})$ denotes the single-particle 
energy and $\mu$ is the chemical potential. As in \cite{Jensen}, we decompose the interaction 
\begin{equation}
\langle{ \textbf{k} \left| V\right|  \textbf{k}'  }\rangle = 4 \pi \sum_l (2 l +1) P_l(\hat{\textbf{k}}\cdot 
\hat{\textbf{k}}') V_l(k,k') 
\end{equation}
and the gap function
\begin{equation}
\Delta(\textbf{k})= \sum_{l m} \sqrt{\frac{4 \pi}{2 l +1}}Y_{l m}(\hat{\textbf{k}})\Delta_{l m}(k) \, ,
\end{equation}
where $Y_{l m}(\hat{\textbf{k}})$ denotes the spherical harmonics, $ l$ is the orbital angular 
momentum, $m$ is its projection along the $z$ axis and $ P_l(\hat{\textbf{k}}\cdot \hat{\textbf{k}}')$ 
refers to the Legendre polynomials. After performing an angle-average approximation (we do 
not retain the $m$-dependence, anisotropic pairing gaps \cite{Khodel} will be discussed in a 
forthcoming paper)  we have the following equation for any value of $l$
\begin{eqnarray}
\label{eq:AAgap}
\Delta^j_l(k)=\sum_{l'}\frac{ (-1)^\Lambda}{\pi}\int{dk'~V^j_{l l'}(k,k')}\frac{\Delta^j_{l'}(k')}{E(k')}{k'}^2 \, ,
\end{eqnarray}
where $\Lambda =1+ (l-l')/2$, $j$ refers to the total angular momentum ($\vec J = \vec l + \vec S$) 
quantum number including spin $\vec S$ and now 
$E(k)^2 =\xi(k)^2+ \sum_{j l}\Delta^j_l(k)^2$. Gaps with different $l$ and $j$ are coupled due to 
the energy denominator but, for the sake of simplicity, we assume that different components of 
the interaction mainly act on non-overlapping intervals in density. This assumption will turn out to 
be correct in the neutron matter case while only partially justified when treating gaps for symmetric 
nuclear matter. To solve Eq.\ (\ref{eq:AAgap}), we follow the approach suggested by Khodel 
{\it et al.} \cite{Khodel S} that has been proven to be stable even for small values of the gap and 
to require only the initial assumption of a scale factor $\delta$ (results, of course, will be 
$\delta$-independent). We define an auxiliary potential $W$ according to
\begin{eqnarray}
W_{l l'}(k,k') = V_{ll'}(k,k') - v_{ll'}\phi_{ll'}(k)\phi_{ll'}(k') \; ,
\end{eqnarray}
where $ \phi_{ll'}(k)=V_{ll'}(k,k_F)/V_{ll'}(k_F,k_F) $ and $v_{ll'}=V_{ll'}(k_F,k_F) $                       
so that $W_{ll'}(k,k')$ vanishes on the Fermi surface.
The coupled gap equations can be rewritten as
\begin{eqnarray}
\Delta_l(k)-\sum_{l'}{ (-1)^\Lambda \int{d\tau'~  W_{ll'}(k,k')\frac{\Delta_{l'}(k')}{E(k')}=
\sum_{l'}{D_{ll'} \phi_{ll'}(k)}   } } \; ,
\end{eqnarray}
where $d\tau=k^2 dk/\pi$ and the coefficients $D_{ll'}$ satisfy
\begin{eqnarray}
\label{eq:coeff}
D_{ll'}=(-1)^\Lambda v_{ll'} \int{d\tau ~\phi_{ll'}(k) \frac{\Delta_{l'}(k)}{E(k)}} \; .
\end{eqnarray}
The gap is defined as follows
\begin{eqnarray}
\label{eq:buildgap}
\Delta_l(k)=\sum_{l_1l_2} D_{l_1l_2}\chi^{l_1l_2}_l(k) \; ,
\end{eqnarray}
where
\begin{eqnarray}
\label{eq:functions}
\chi^{l_1l_2}_l(k)-\sum_{l'}{(-1)^{\Lambda} \int   {d\tau'~ W_{ll' }(k,k') \frac{ \chi^{l_1l_2}_{l'}(k')}{E(k')} } }
=\delta_{ll_1}  \phi_{l_1l_2}(k) \; ,
\end{eqnarray}
and $\delta_{ll'}$ is the scale factor.
The property that $W_{ll'}(k,k') $ vanishes on the Fermi surface ensures a very weak 
dependence of $ \chi^{l_1l_2}_l(k)$ on the exact value of the gap so that, in first approximation, 
it is possible to rewrite the previous equation (\ref{eq:functions}) as
\begin{eqnarray}
\label{eq:functionsFO}
\chi^{l_1l_2}_l(k)-\sum_{l'}{(-1)^{\Lambda} \int   {d\tau'~ W_{ll' }(k,k') \frac{ \chi^{l_1l_2}_{l'}(k')}{\sqrt{{\xi^2(k')}+\delta^2}} } }
=\delta_{ll_1}  \phi_{l_1l_2}(k) \; .
\end{eqnarray}
We use this equation to evaluate $\chi^{l_1l_2}_l(k)$ initially by matrix inversion, then we use 
this function to self-consistently evaluate $D_{ll'}$. Finally, we solve the system given by 
Eqs.\ (\ref{eq:coeff})--(\ref{eq:functions}) in a self-consistent procedure.
We always assume $\mu = \varepsilon_F$ and adopt the relativistic version of the single-particle 
energy $\varepsilon \left( k \right) = \sqrt{k^2 + M_N^2 }$, where $M_N$ is the nucleon mass.
In principle, the effective force to be included in Eq.\ (\ref{gapeq}) should be generated by the sum
of all particle-particle irreducible Feynman diagrams~\cite{Mig.67}, but
in most applications to nuclear systems only the bare nucleon-nucleon 
interaction is kept~\cite{Jensen}. Corrections to the bare force, caused by
medium polarization effects (see Refs.\ \cite{Sch.96, Lombardo:2005sw, Baldo:2010du} and 
references therein) will be neglected in the present analysis and postponed to a forthcoming 
paper. As a consequence, for the pairing potential $V(p,k)$ we introduce the following ansatz:
\begin{equation}
\label{V_pk}
V(p,k) = V_{2B} (p,k) + V_{3B}(p,k,m) \simeq V_{2B} (p,k) + V^{eff}_{2B} (k_F,p,k) \; ,
\end{equation}
where $V_{2B}$ is the Idaho \cite{Machleidt:2011zz} NN potential at N3LO in the chiral 
expansion\footnote{Among the different versions, we employ the chiral potential in 
which the regulator function $f(p{'},p) = \exp[-(p'/\Lambda)^{2n} - (p/\Lambda)^{2n}]$ has
the cutoff $\Lambda = 500$\,MeV, and $n=2$ for the $2\pi$ exchange 
contributions \cite{n3lo_2003}.}
or the Juelich version \cite{Epelbaum:2008ga}, and
the three-body potential is approximated by an effective two-body density-dependent
potential $V^{eff}_{2B}$ derived by Holt {\it et al.} in Refs.\ \cite{Holt:2009,Holt:2010}. 
We employ in our calculations the evolved two-body potentials $V_{{\rm lowk}}$ 
(with a smooth cut-off in momentum space \cite{smooth}) and $V_{{\rm srg}}$ \cite{rengroup} 
using two different evolution operators (see Sect.\ \ref{hpw_sec} for more details).
When considering self-energy effects, we simply perform the transformation $ M_N 
\rightarrow M_N^*$ using the effective mass obtained by Holt {\it et al.} in Ref. \cite{meff1} 
using a density matrix expansion technique. In Ref.\ \cite{meff1} the two-body interaction 
was comprised of long-range one- and two-pion exchange contributions and a set of contact terms 
contributing up to fourth power in momenta as well as the leading order chiral three-nucleon 
interaction. The explicit formula is given by
\begin{equation}
\label{effmass}
M^*(\rho) = M \left(1+2 M F_\tau(\rho) -\frac{k_F^2}{2 M^2}   \right)^{-1}\,,
\end{equation}
where the strength function $F_\tau(\rho)$ is defined as follows
\begin{equation}
F_\tau(\rho) = \frac{1}{2 k_F}\left(\frac{\partial U(p,k_F)}{\partial p}\right)_{p=k_F}
\end{equation}
with $U(p,k_F)$ the single particle potential and $-\frac{k_F^2}{2 M^2}$ a relativistic 
correction. 
In Fig.\ \ref{fig:meff} we plot the effective masses for nuclear and neutron matter as 
functions of density. From the effective mass behavior we can expect that
the self-energy effects will play a central role in the high-density components of the gap, 
while at low densities the effects will be rather negligible. Second-order perturbative contributions 
to the single-particle energies are expected to increase the effective mass \cite{Holt13}.

%
%

\section{Pairing gaps}

\subsection{Singlet channel ($^1S_0$)}
\label{1s0_sec}

\begin{figure}
\begin{center}
\includegraphics[width = 12.5 cm]{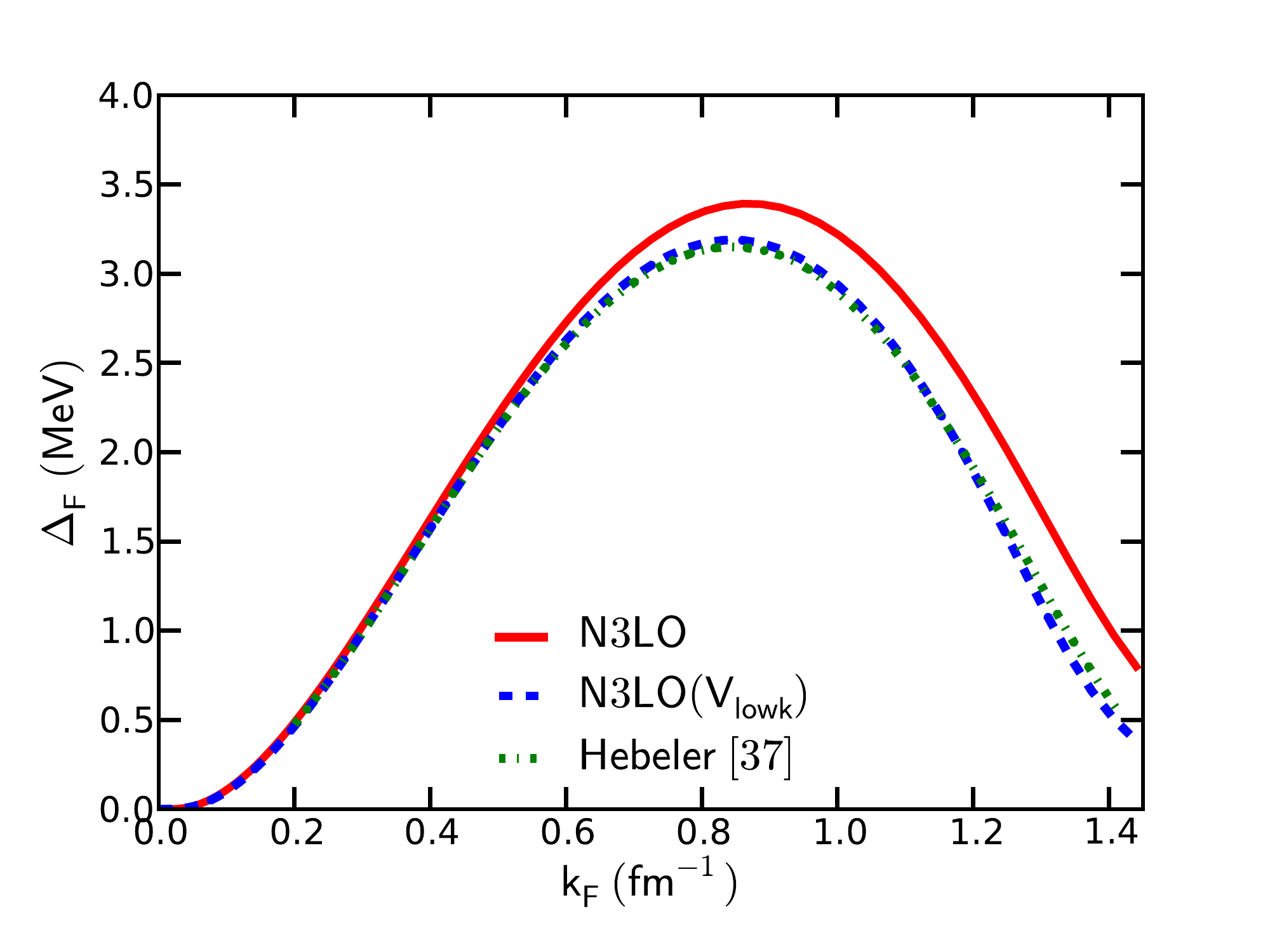}
\caption{(color online) The $^1S_0 $ gap for nuclear matter computed with the realistic chiral potential of Ref.\ 
\cite{Machleidt:2011zz, n3lo_2003} at N3LO (red line) and the corresponding 
$V_{{\rm lowk}}$ potential (blue dashed line). With the green dashed-dotted line we include, as a 
benchmark, a similar calculation performed by Hebeler {\it et al.} \cite{Vlowbcs}.}\label{fig:1s0npBCS}
\end{center}
\end{figure}

In the singlet channel, the only difference between the $nn$ and 
$np$ potential is the charge independence breaking and the charge symmetry breaking terms,
which are treated as small perturbations in ChPT. In Fig.\ \ref{fig:1s0npBCS} we test our solution
to the gap equation, in the nuclear matter case, against previously published results \cite{Vlowbcs} 
with the low-momentum interaction $V_{{\rm lowk}}$. In addition, we compute the $^1S_0$ gap
from the bare chiral NN interaction and find a qualitatively very similar behavior. The gap 
$\Delta_F$ reaches a maximum value of approximately 
3.5 MeV at $k_F \simeq 0.85$\,fm$^{-1}$ when the bare interaction is used in the two-body sector, 
while a somewhat reduced gap (by almost $0.5$ MeV) if we consider $V_{{\rm lowk}}$. 

\begin{figure}
\begin{center}
\includegraphics[width = 12.5 cm]{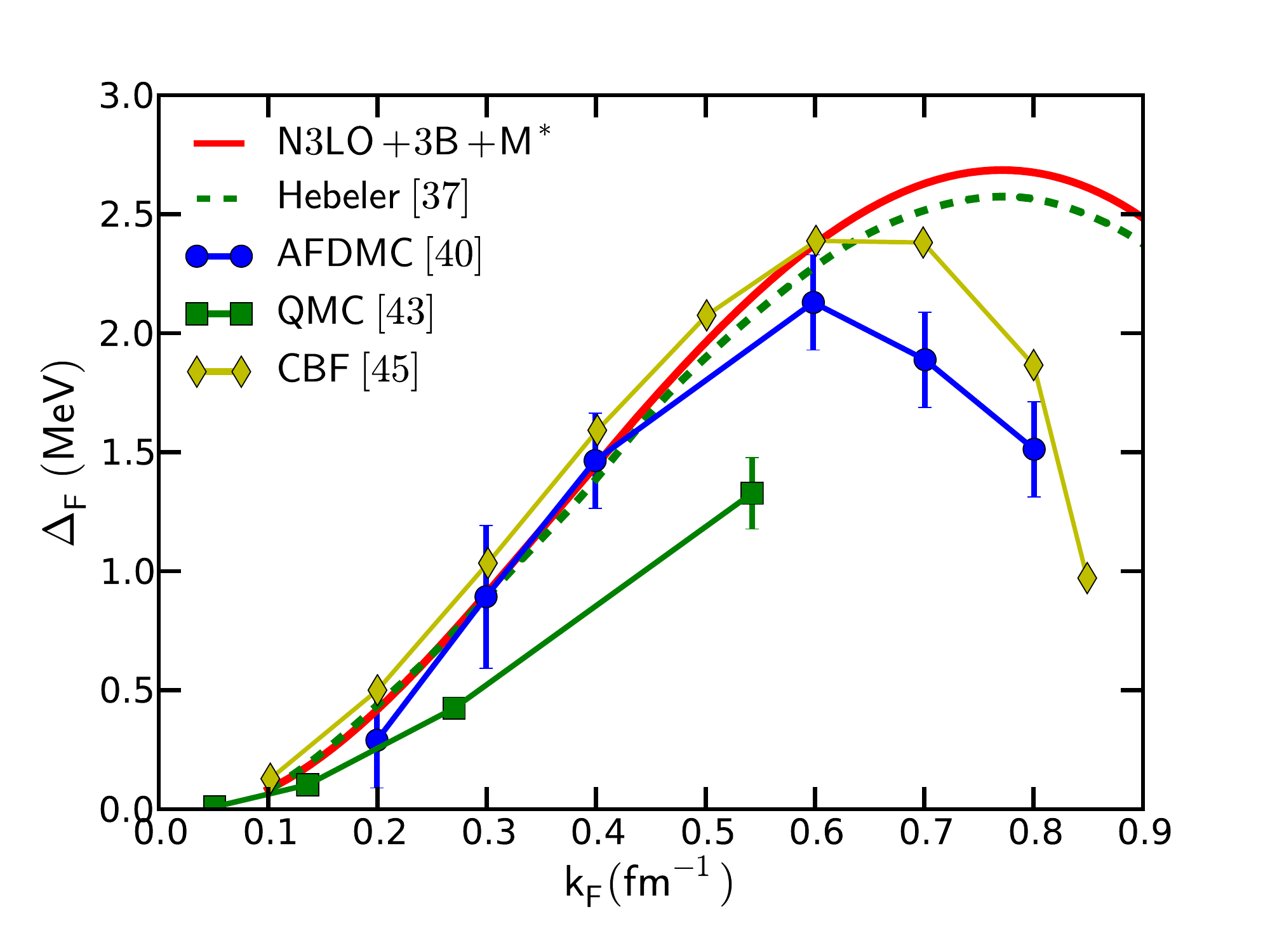}
\caption{(color online) The $^1S_0 $ gap for neutron matter computed with the realistic chiral potential of 
\cite{n3lo_2003,Machleidt:2011zz} at N3LO plus the three-body contribution of Eq.\
(\ref{V_pk}) and the inclusion of the effective mass in Eq.\ (\ref{effmass}). As a comparison, we include 
a similar calculation by Hebeler \cite{Vlowbcs} with a green dashed line and a set of {\it ab-initio} simulations with 
different many-body techniques: AFDMC (blue circles) \cite{Gandolfi:2009tv}, QMC Green Functions (green squares) \cite{Gezerlis} and CBF (yellow diamonds) \cite{cbf}. See the text for additional details.
}\label{fig:1s0nnBCS}
\end{center}
\end{figure}

In the neutron matter case, at the two-body level, there is good agreement with the gap computed from well known realistic potentials like the CD-Bonn or Nijmegen interactions \cite{Jensen, CDBONN, Nijmegen}, but at larger densities the N3LO gap exhibits a higher value.
This can be explained by observing that the phase shifts from the chiral N3LO potential
exhibit more attraction than the CD-Bonn potential for high momenta, as already observed 
by Hebeler \textit{et al.} \cite{Vlowbcs}. In Fig.\ \ref{fig:1s0nnBCS} we compare our full 
calculation for the gap, i.e., with the complete potential in Eq.\ (\ref{V_pk}) and the density-dependent 
effective mass in Eq.\ ({\ref{effmass}), with recent results by Hebeler {\it et al.} \cite{Vlowbcs}, where 
the authors started from a chiral N3LO interaction and evolved to a sharp
low-momentum interaction\footnote{We used a rather different approach to construct our 
$V_{{\rm lowk}}$. The RG procedure has been performed with different cutoffs and 
regulating functions, in particular a Fermi-Dirac function $ f_\Lambda(k)=1/({1+e^{(k^2 - 
\Lambda^2)/\varepsilon^2}})$ and an exponential cutoff $ f_\Lambda(k)=e^{-(k^2/\Lambda^2)^n}$
\cite{smooth}. The results show a very weak cutoff-dependence.}. Also presented for comparison 
are {\it ab-initio} 
results obtained in the last several years: Auxiliary Field Diffusion Monte Carlo (AFDMC) 
\cite{Gandolfi:2009tv} with AV8' \cite{av8} + UIX \cite{uix} potentials, Quantum Monte Carlo (QMC)
\cite{Gezerlis}, where the authors have retained the $S$-wave part of the AV18 \cite{av18} 
interaction, and Correlated Basis Functions (CBF) \cite{cbf} still with AV8' plus UIX. We 
observe that at low densities the gap behaviors are very similar, but beyond Fermi momenta 
of $k_F \approx 0.6 \mbox{ fm}^{-1}$ the gaps computed with the Argonne potentials 
decrease rapidly in contrast to those from chiral interactions. At the present time, it is hard 
to assess if disagreement is due to different choices in the nuclear Hamiltonian or different
many-body methods.

It is useful to consider separately the different physical effects governing the $^1 S_0$ pairing gap. 
In Fig.\ \ref{fig:1s0nnBCS_contr} we plot the gaps obtained with two-body interactions alone
(the dotted lines represent the bare and the renormalized N3LO potentials), 
with the inclusion of effective three-body forces (dashed lines) and considering also 
self-energy effects (solid lines).
\begin{figure}
\begin{center}
\includegraphics[width = 12.5 cm]{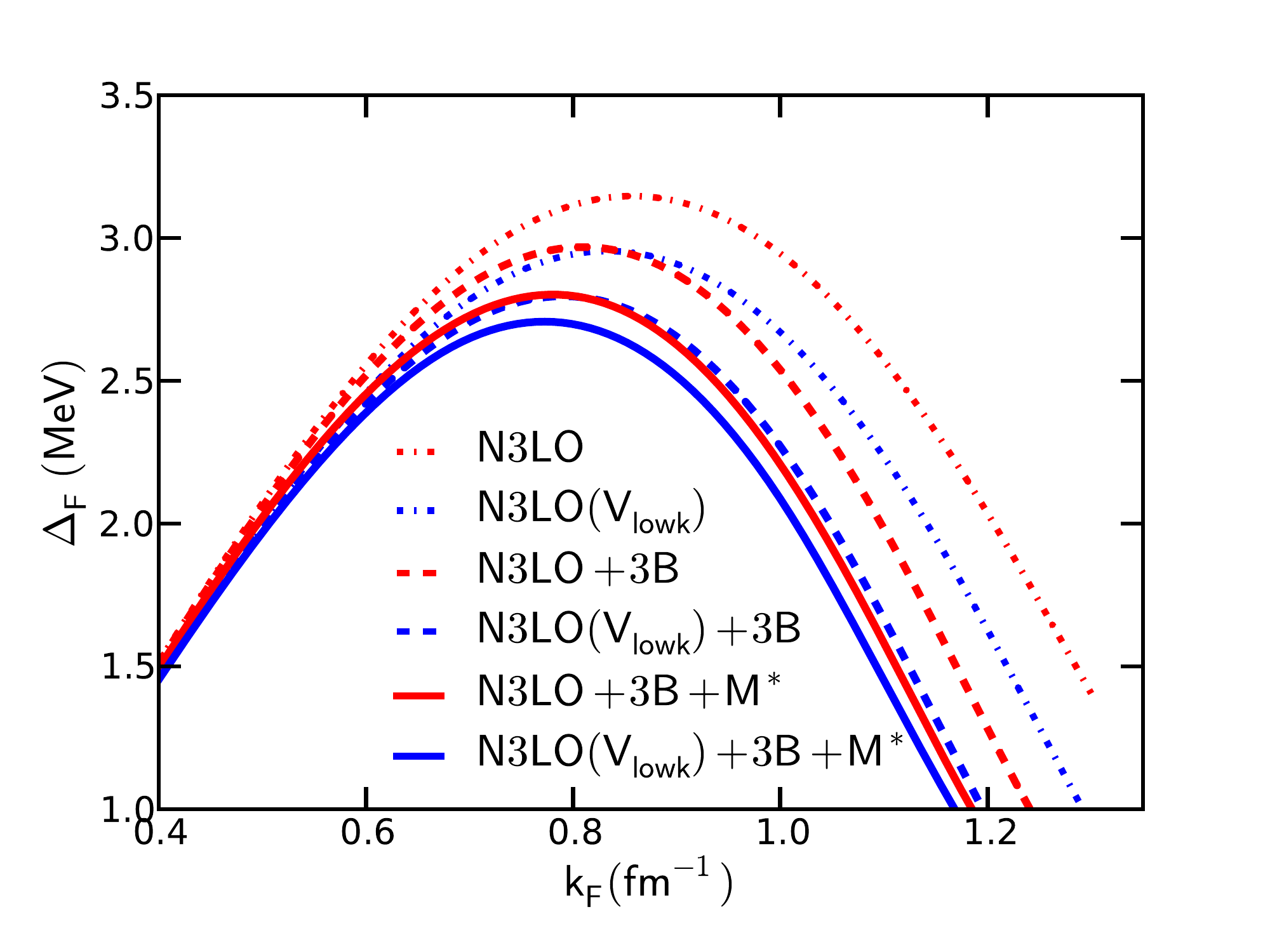}
\caption{(color online) The $^1S_0 $ gap for neutron matter. In this figure we show all the contributions
to the pairing gap $\Delta_F$ starting from the inclusion of the bare two-body potential 
(red dotted line) or $V_{{\rm lowk}}$ (blue dotted line) and then including effective three-body forces 
(dashed lines) and a density-dependent effective mass (solid lines).}\label{fig:1s0nnBCS_contr}
\end{center}
\end{figure}
By construction, we expect that at low densities the three-body effects are rather small, while 
only at higher densities do they become appreciable. The main role of both three-body forces 
and the effective mass is to substantially reduce the attractive strength in the $S$ channel
(for higher partial waves the situation is more involved, see Sect.\ \ref{hpw_sec}).

\subsection{Higher partial waves ($^3SD_1$ and $^3PF_2$)}
\label{hpw_sec}

In addition to the $^1S_0$ channel, in the nuclear matter case a non-vanishing gap appears in
the $ ^3 SD_1 $ channel. The presence of a bound state in this channel and the very high phase 
shifts in the $^3 S_1$ channel indicate that the interaction is more attractive than 
in the other channels. As a consequence the gap has a magnitude of about 10 MeV, as can be seen 
in Fig.\ \ref{fig:gap3s1bcs}, with conventional realistic potentials.
There is no agreement on the details of the gap in this channel, but both Elgar{\o}y {\it et al.} \cite{Elg} 
and Takatsuka {\it et al.}  \cite{TaTa} suggest the possibility of a gap of such magnitude (see curves labeled, respectively, by BONN-A and OPEG in Fig.\ \ref{fig:gap3s1bcs}).
While BONN-A \cite{BonnA} is a complete one-boson exchange potential, OPEG \cite{TaTa} contains 
only the one-pion exchange tail and a Gaussian repulsive core.
\begin{figure}
\begin{center}
\includegraphics[width= 12.5 cm]{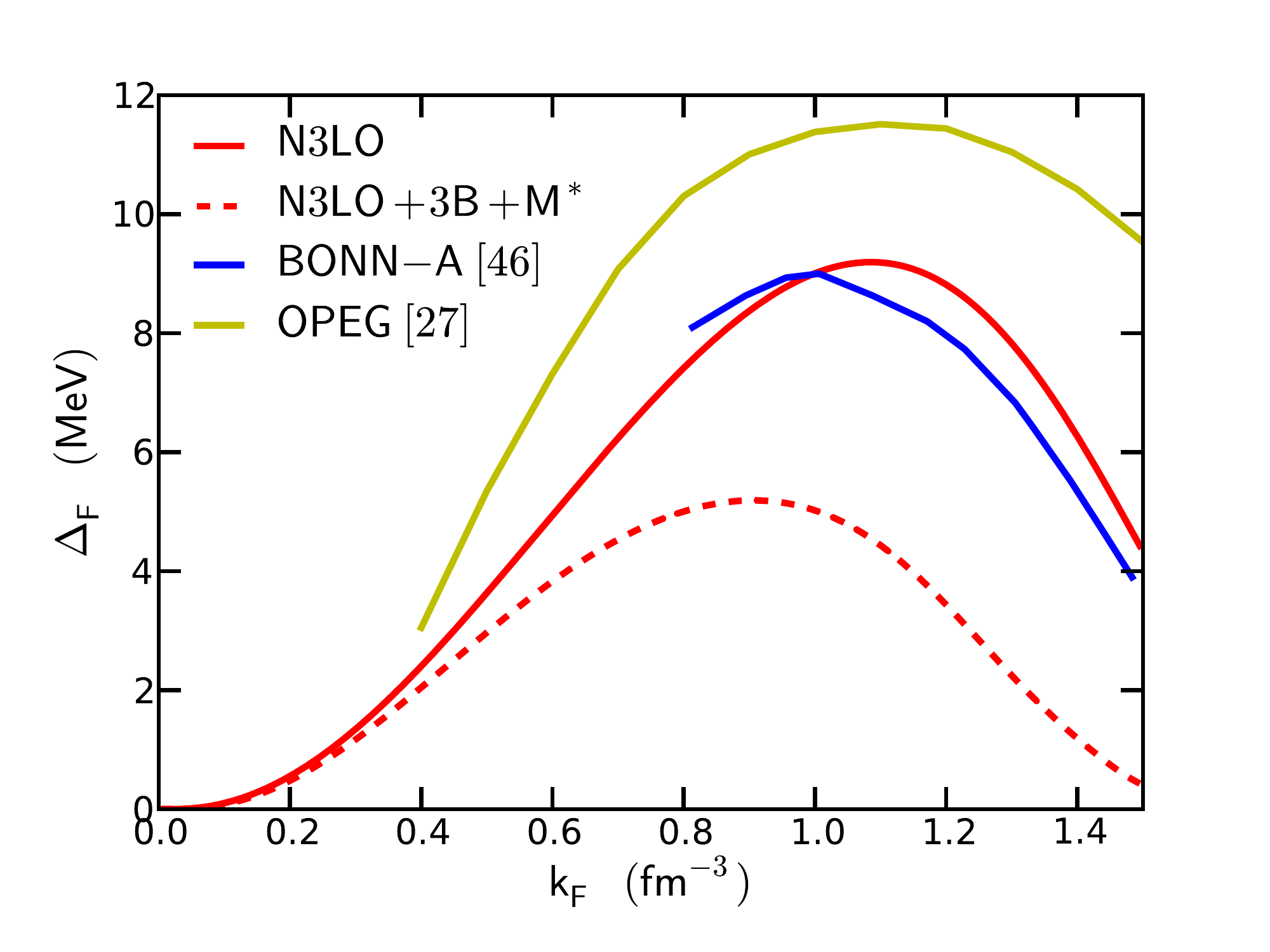}
\caption{The gap in the $^3SD_1$ channel. We plot our calculations with the N3LO interaction 
(red line) in comparison with results obtained employing BONN-A potential \cite{Elg} 
(blue curve) and OPEG \cite{TaTa} (yellow line). All results suggest a very large pairing 
gap (around 10 MeV), but complete calculations including three-body forces and effective
masses (see Eqs.\ (\ref{V_pk}) and (\ref{effmass})), shown in the dashed red
curve, indicate a substantial reduction and a sizable modification of the gap's shape.}
\label{fig:gap3s1bcs}
\end{center}
\end{figure}
The combined effect of three-body forces and self-energy effects leads not only to a 
sizable reduction of the gap itself but also to a shift 
of the gap's maximum at $k_F \approx 1 
\mbox{ fm}^{-1}$ and a rapid decrease at higher Fermi momenta.
\begin{figure}
\begin{center}
\includegraphics[width= 12.5 cm]{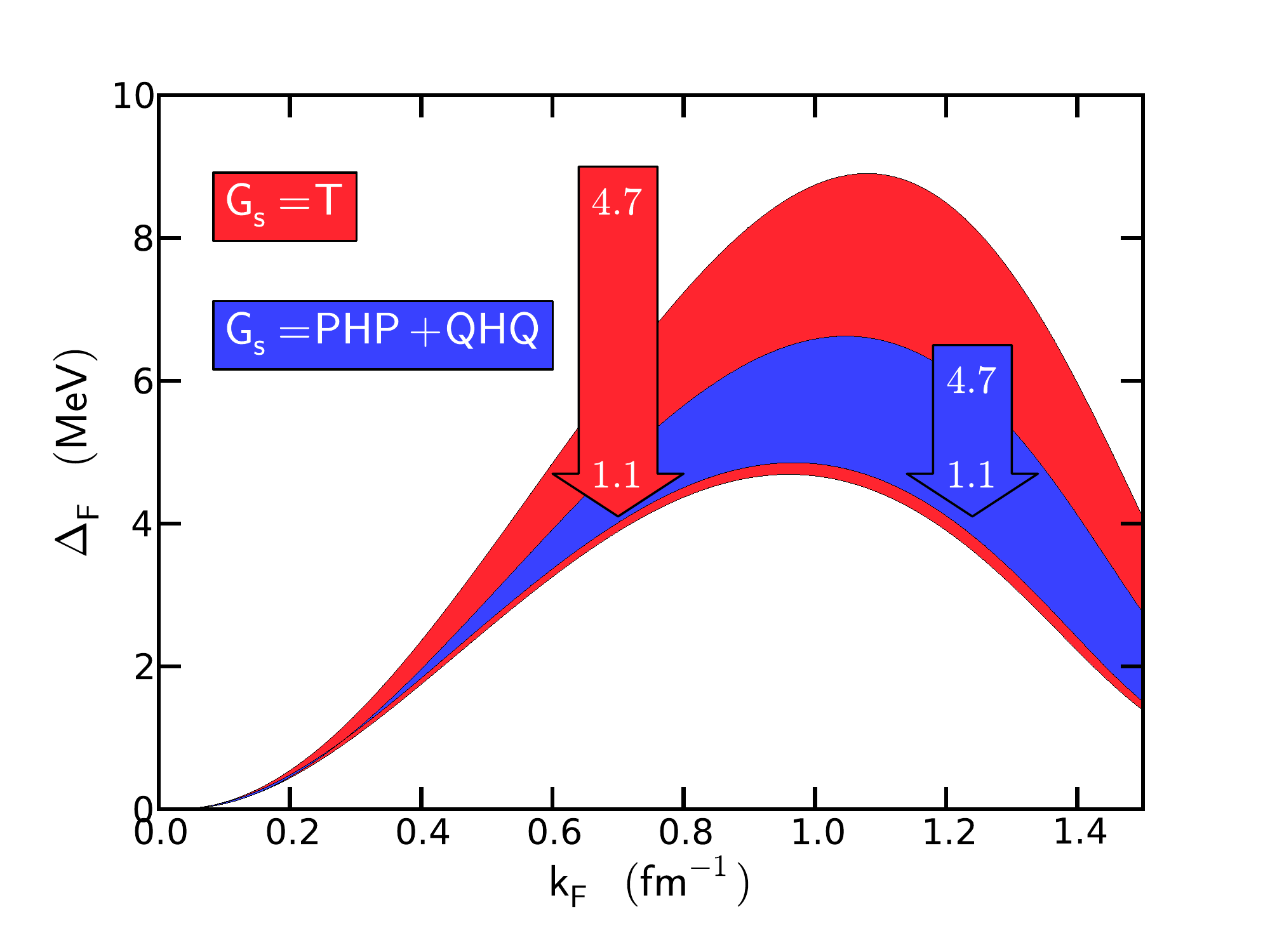}
\caption{(color online) The evolution of the pairing gap in the $^3SD_1$ channel with 
SRG-evolved interactions. We employed two different
evolution operators $G_s$ : $T_{rel}$ (red band) and $PHP+QHQ$ 
(blue band). The arrows denote the flow variable $\lambda$
 (related to $s$ through $\lambda \equiv s^{-1/4}$) which 
is varied from 4.7 fm $^{-1}$ down to 1.1 fm$^{-1}$.}
\label{fig:gap3s1bcs2}
\end{center}
\end{figure}

Due to the large densities over which the pairing gap remains finite, it is questionable whether 
low-momentum interactions, $V_{{\rm lowk}}$, with a block-diagonal momentum-space cutoff on
the order of $\Lambda \sim 2.0$\,fm$^{-1}$ are appropriate.
A better approach is provided by the Similarity Renormalization Group (SRG), where 
off-diagonal momentum-space matrix elements are supressed.
In this case, we study nuclear Hamiltonians $H=T_{rel} + V$ evolved through the 
SRG procedure \cite{rengroup}, where we define a class of Hamiltonians
\begin{equation} 
H_s= U_s H U^\dag_s \equiv T_{rel} + V_s
\end{equation} with a generator 
\begin{equation}
\eta_s=\frac{dU_s}{ds}U_s^\dag=-\eta_s^\dag \; .
\end{equation}
If we choose $\eta_s=\left[G_s,H_s\right]$  the flow equation takes the form
\begin{equation}
\label{eq:SRGflow}
 \frac{dH_s}{ds}=\left[\left[G_s,H_s\right], H_s\right] \; .
\end{equation}
As shown in \cite{rengroup}, results obtained from SRG-evolved interactions are very similar 
to those obtained from $V_{\rm lowk}$ 
if an appropriate $G_s$ is chosen. Moreover, the SRG interaction has many salient 
features of low-momentum interactions, such as independence of the physical observables 
from the operator $G_s$, perturbativeness and universality.  In the literature it is common to 
encounter also the dimensional parameter $\lambda = s^{-1/4}$ fm$^{-1}$.
A very interesting feature of the SRG procedure is that the tensor interaction strength is 
reduced as $s$ increases, and this modification to the interaction can strongly modify the 
$^3SD_1$ gap. Since all physical observables should remain unchanged under an SRG 
transformation, this variation represents an uncertainty estimate in the pairing strength.
A common choice for $G_s$ is $T_{rel}$, and in this case as $s$ increases, $V_s$ approaches 
the diagonal form.
\begin{figure}
\begin{center}
\includegraphics[width=12.5 cm]{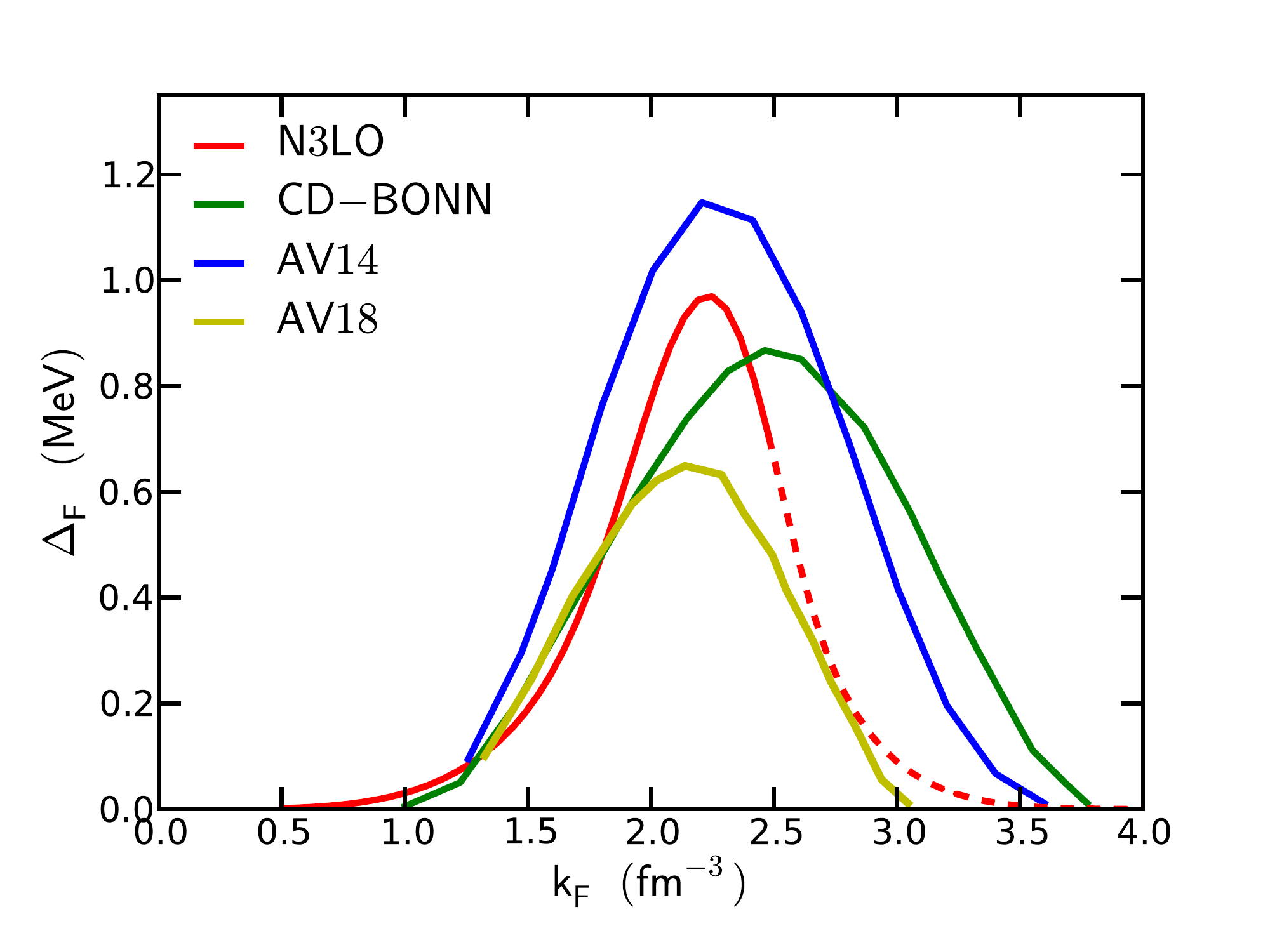}
\caption{(color online) The gap in the $^3PF_2 $ channel obtained from the N3LO \cite{n3lo_2003} 
(red line) interaction in comparison with several realistic NN potentials taken from Ref. \cite{Jensen}. 
Chiral potentials, by definition, can be trusted only up to momenta close to the cutoff 
(beyond the cutoff, the pairing gap is symbolized with a dashed line).}
\label{fig:gap3p23f2bcs}
\end{center}
\end{figure}
We tested one more generator
\begin{equation}
\label{eq:PQ}
G_s =P_\Lambda H_s P_\Lambda + Q_\Lambda H_s Q_\Lambda \;,
\end{equation}
where $P_\Lambda$ and $Q_\Lambda$ are, respectively, the projector and the exclusion operators 
in the subspace $\left\{ k<\Lambda \right\}$ (see Sect. 3.4 in Ref. \cite{rengroup}).
From Eq.\ (\ref{eq:SRGflow}) it is easy to see that, if $H$ 
is a two body Hamiltonian expressed in the second quantization formalism,
$ \left( dH_s/ds \right)_{s=0}$ will also include three body interactions. In this way,
the evolution over the flow will naturally induce many-body interactions.
The errors arising from omitting the induced many-body forces can be estimated by analyzing the 
dependence of the physical observables on the flow parameter $\lambda$.
Our results are shown in Fig.\ \ref{fig:gap3s1bcs2}, where we tested the two evolution operators.
For $G_s = T_{rel}$ (red color) we found that the gap becomes quite stable for 
$\lambda < 2.2$\, fm$^{-1}$, where the maximum is reduced to approximately 5 MeV 
(a factor of 2 smaller compared to the bare potential). In the range $1.3 \mbox{ fm}^{-1} \leq 
\lambda \leq 2.2 \mbox{ fm}^{-1}$ the variation in the size of the gap is on the order of 0.5 MeV or
less. When using $G_s$ given by Eq.\ (\ref{eq:PQ}) we obtained very similar results, confirming the approximate independence of the 
physical results on the choice of $G_s$, but with a reduced cutoff-dependence.
\begin{figure}
\begin{center}
\includegraphics[width=12.5 cm]{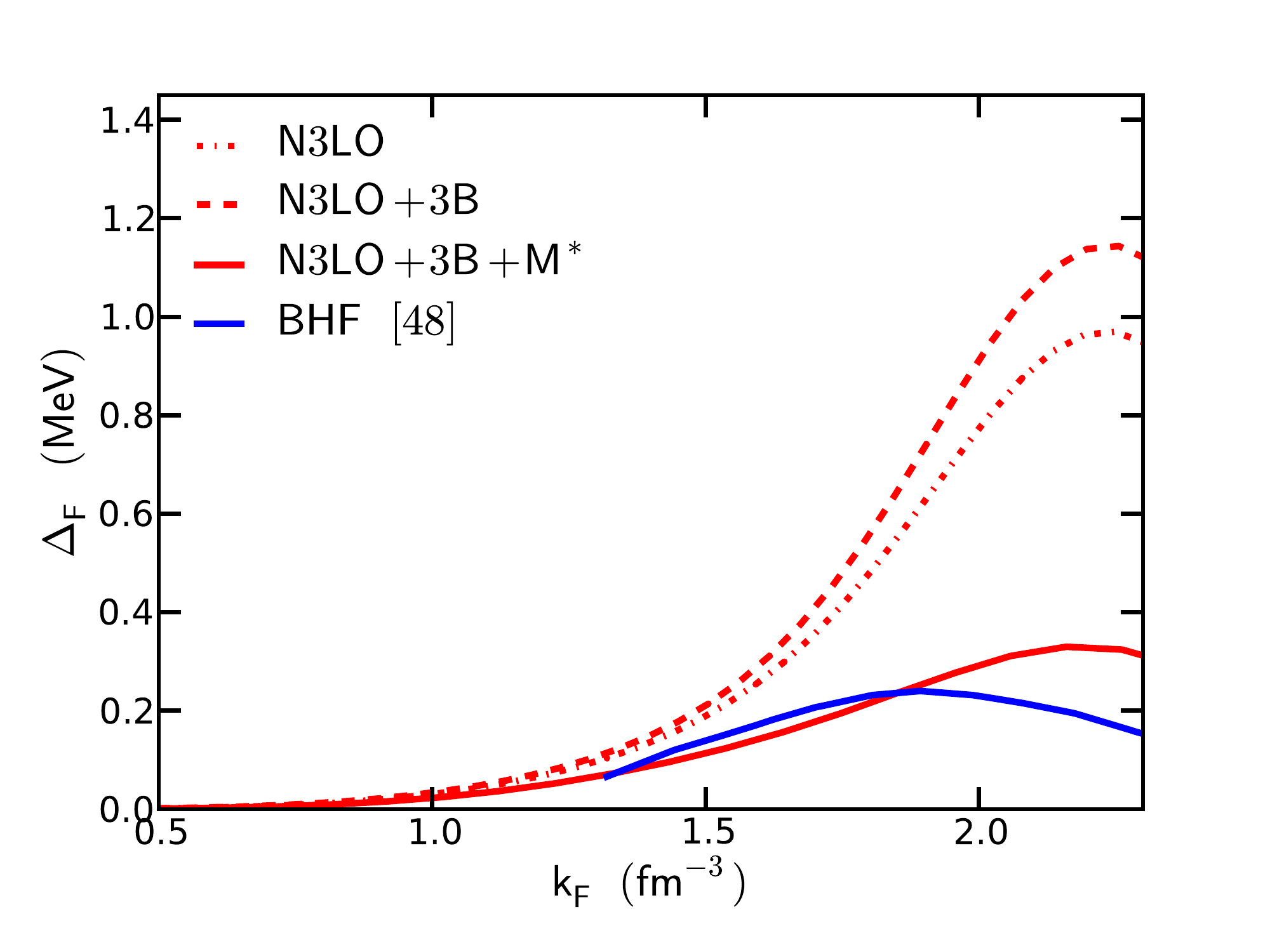}
\caption{(color online) The gap in the $^3PF_2 $ channel with only N3LO potential (dotted red line), with 
three-body forces (dashed red line) and including also self-energy effects (solid red line). In 
comparison we plot the results of recent BHF calculations \cite{PhysRevC.87.062801} (blue curve).}
\label{fig:gap3p23f2bcs_2}
\end{center}
\end{figure}

In the neutron matter case, while at low density the dominant channel is the 
$ ^1 S_0 $ partial wave, at higher densities the high-momentum components (which are repulsive)
become more important, suppressing the gap, and this happens at $ k_F \approx 1.5$\, fm$^{-1}$.
At these densities, the only channel which substantially contributes to the neutron matter gap is 
the coupled $ ^3 PF_2 $, where the coupling is due to the tensor interaction.
As can be seen in Fig.\ \ref{fig:gap3p23f2bcs}, there is a significant dependence of the gap on the 
potential model, though the peak in the gap consistently occurs between $2.2 \leq k_F 
\leq 2.6$\, fm$^{-1}$. At the high densities and associated momentum scales relevant for 
pairing in this channel, realistic NN interactions are not as well constained by fits to phase 
shifts, which partially explains the differences in the observed gaps.
As explained in \cite{Machleidt:2011zz}, in this channel one expects a crucial contribution from the
three-pion-exchange topology at N$^4$LO and from the contact term at N$^5$LO,
which should reduce the attraction in this channel.
All reasonable interactions give a gap of magnitude $\approx 1$ MeV, and we expect a small but not 
negligible reduction of the gap from the higher orders in $Q/\Lambda_\chi$.
In Fig.\ \ref{fig:gap3p23f2bcs_2} we plot predictions for the $ ^3 PF_2 $ gap including three-body forces 
(dashed red line) and self-energy effects (solid red line) in 
comparison with a very recent Brueckner-Hartree-Fock
calculation by Dong {\it et al.} \cite{PhysRevC.87.062801}, where the authors employed the Bonn B 
potential \cite{Bonnb} and a microscopic three-body force constructed by Li {\it et al.} \cite{li}. 
Our complete calculation nicely agrees with \cite{PhysRevC.87.062801}, in particular for 
small momenta, and suggests a sizeable reduction of the gap if many-body forces are taken into account.

In Fig.\ \ref{fig:gap3p23f2bcs_3} we show also the $^3PF_2$ gap we have computed from the Juelich 
ChPT potentials \cite{Epelbaum:2008ga}. 
Because the $^3PF_2$ gap extends towards very large densities (even beyond the reasonable limits of applicability of a ChPT approach) is very interesting to test the robustness of previous calculations 
(see Figs. \ref{fig:gap3p23f2bcs} and \ref{fig:gap3p23f2bcs_2}) against a different theoretical approach. In fact, in the last years Epelbaum {\it et al.} developed a new scheme in the construction of a realistic chiral potential where, instead of a Dimensional Regularisation scheme for chiral-loop integrals, a finite cutoff
$\Lambda$ is kept in the range of $500-800$ MeV which appears to be physically reasonable  and matches well with the cutoff used in the Lippmann-Schwinger (LS) equation.
As a consequence, in our calculations we employed two different cutoffs: $\Lambda_{LS}$ for the LS equation (with non relativistic kinematics) and $\Lambda_{2\pi}$ for the spectral-function regulator (SFR) of 
the two-pion exchange potential (varied between $500$ and $700$ MeV). For Fermi momenta up to 
nearly $k_F = 1.4$\, fm$^{-1}$, the predictions from the different potentials are nearly universal
and agree reasonably well with the predictions from the Entem and Machleidt chiral N3LO potential. 
However, beyond this density there is a significant scale dependence in the theoretical predictions, in particular to $\Lambda_{2\pi}$.  This uncertainty has to be taken into account if microscopic $^3PF_2$ gaps are used to describe the cooling process of neutron stars \cite{futurepaper}}.

\begin{figure}
\begin{center}
\includegraphics[width=12.5 cm]{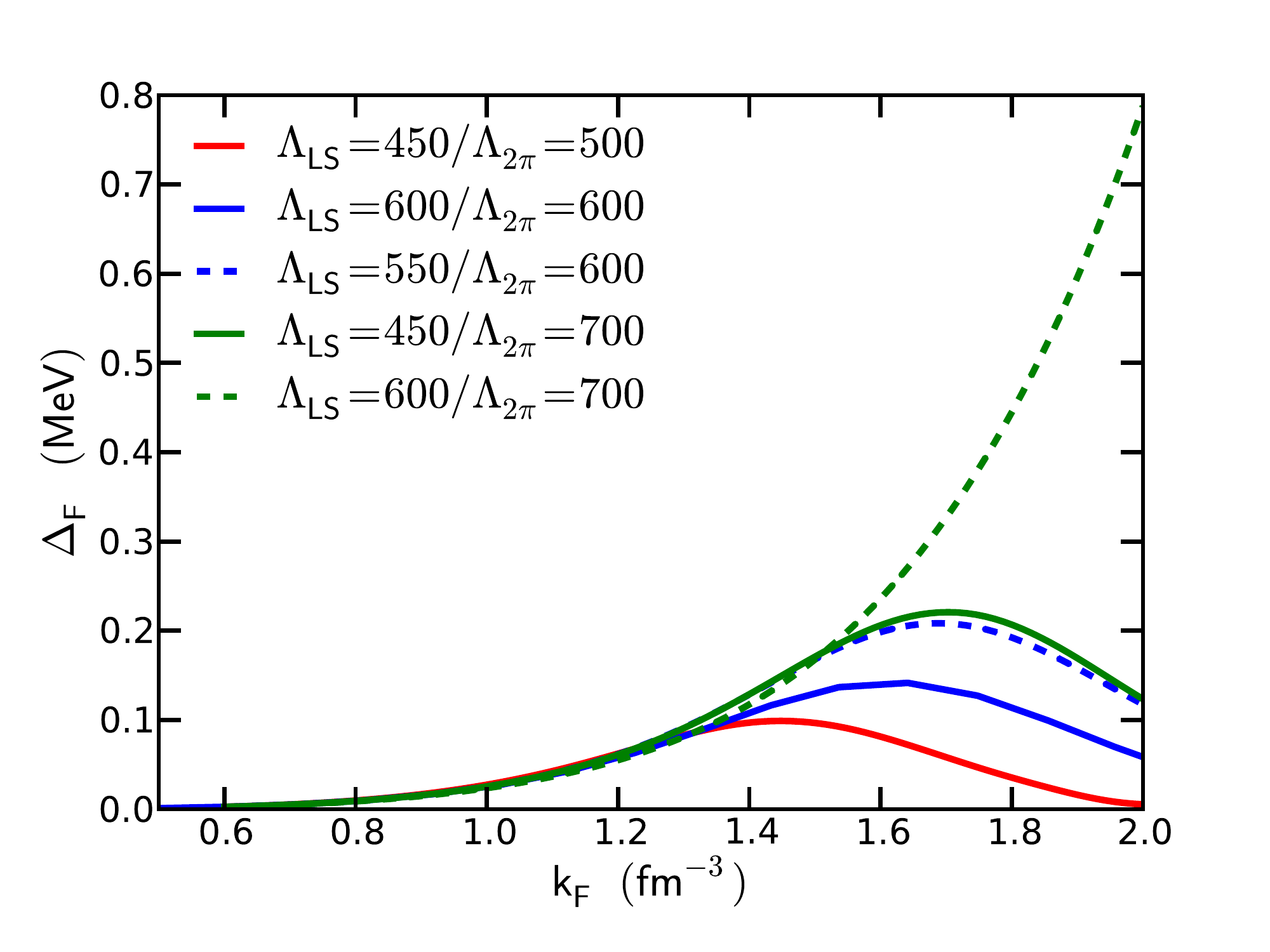}
\caption{The gap in the $^3PF_2 $ channel as a function of the resolution scale in the Juelich 
nucleon-nucleon interactions \cite{Epelbaum:2008ga}. 
The scales refer respectively to the cutoffs (units of MeV) in the 
Lippmann-Schwinger equation ($\Lambda_{LS}$) and the spectral function regulator in multi-pion exchange loop diagrams($\Lambda_{2\pi}$). It appears that the magnitude of the gap's maximum 
is very sensitive to $\Lambda_{2\pi}$ and, to a lesser content, to $\Lambda_{LS}$.}
\label{fig:gap3p23f2bcs_3}
\end{center}
\end{figure}

\section{Conclusions}
We have presented calculations of the pairing gaps in infinite nuclear and neutron matter
employing realistic two- and three-body nuclear forces derived within the framework of chiral 
effective field theory. The BCS gap equation is solved employing Khodel's method, which is found 
to be stable even for small values of the pairing gap. Three-nucleon forces help reduce the 
strength of pairing in the $^1S_0$ and coupled $^3SD_1$ channels, while for the coupled 
$^3PF_2$ channel the three-nucleon forces enhances the gap. In all cases considered in the present
work, consistent nucleon effective masses reduce pairing correlations. Of particular interest is the
scale dependence of the $^3PF_2$ pairing gap, which exhibits a nearly universal behavior at 
low densities in all chiral potentials considered. This works sets the stage for future applications 
to pairing gaps in finite-temperature neutron matter \cite{futurepaper}.

\label{secIV}

\section{Acknowledgements}
Work supported in part by US DOE Grant No.\ DE-FG02-97ER-41014. The authors are deeply 
grateful to E. Epelbaum (Institut f\"ur Theoretische Physik II,
Ruhr-Universit\"at Bochum) for providing the chiral potential of Ref. \cite{Epelbaum:2008ga}.

\end{document}